\documentclass[thmsa,12pt]{article}
\normalfont\sffamily
\usepackage{amsfonts}
\usepackage{amssymb}
\usepackage[active]{srcltx}
\usepackage{amsmath}
\usepackage[amsmath,thmmarks,hyperref]{ntheorem}
\usepackage{graphicx}
\usepackage{textcomp}
\usepackage{a4wide}
\usepackage{chicago}
\usepackage{color}
\usepackage[normalem]{ulem}
\usepackage{comment}
\usepackage{pxfonts}
\usepackage{titlesec}
\usepackage{hyperref}

\usepackage{setspace}

\usepackage[english]{babel}

\newtheorem{theorem}{Theorem}
\newtheorem{proposition}{Proposition}
\newtheorem{lemma}{Lemma}
\newtheorem{claim}{Claim}

\newcommand{\SG}{\mathcal{S}_n}

\newcommand{\R}{\mathcal{R}}

\def\barS{\overline{S}}

\usepackage{hyperref}

\begin{document}
	
	\def\title #1{\begin{center}
			{\Large {\sc #1}}
	\end{center}}
	\def\author #1{\begin{center} {#1}
	\end{center}}
	\def\address #1{\vspace{-2em}\begin{center}\small {#1}
	\end{center}}

	
		\title{\sc The roll call interpretation of the Shapley value 
		}
		
		\bigskip
		
		\author{Sascha Kurz}\address{\small Dept.\ of Mathematics, University of Bayreuth\\ Tel: +49 921 557353, e-mail: sascha.kurz@uni-bayreuth.de}

		\author{Stefan Napel}\address{Dept.\ of Economics, University of Bayreuth}

		\begin{center} {\tt \today} \end{center}
		
		\vspace{0.01cm}
		
		\begin{center}{\bf {\sc Abstract}} \end{center}
		{\small
		The Shapley value is commonly illustrated by roll call votes in which players support or reject a proposal in sequence. If all sequences are equiprobable, a voter's Shapley value can be interpreted as the probability of being pivotal, i.e., to bring about the required majority or to make this impossible for others. We characterize the joint probability distributions over cooperation patterns that permit this roll call interpretation: individual votes may be interdependent but must be exchangeable.
		}
		\vspace{0.2cm}
		
		\begin{description}
			{\small
				\item[Keywords:]
				Shapley value; Shapley-Shubik index; roll call model; voting power
				\item[JEL codes:]
				C71; 
				D70; 
				D72  
			}
		\end{description}
		
	
	
	
\section{Introduction}
\label{sec:intro}
\noindent
A player's \emph{Shapley value} equals its expected contribution to surplus creation if 
full cooperation among players is established in random order.
Going back to \citeN{Shapley:1953} and \citeN{Shapley/Shubik:1954}, this is often illustrated by voting games: shareholders, delegates to a council, parties, etc.\
cast their respective voting weight in favor of a proposal one after another.
If player~$i$'s vote is the first to reach the required majority threshold, it `swings' the status of the coalition $S$ of earlier supporters from losing ($v(S)=0$) to winning ($v(S\cup\{i\})=1$); $i$ is then attributed a `marginal contribution' of $v(S\cup\{i\})-v(S)=1$.
Averaging these contributions across all equiprobable voting sequences yields $i$'s Shapley value $\varphi_i(v)$.
It is equal to the probability that $i$ is decisive for passing a proposal. This is commonly interpreted as voting power and also called $i$'s \emph{Shapley-Shubik index (SSI)}.

The implicit assumption in this well-known roll call interpretation of Shapley value and SSI is that all voters support the proposal, i.e., every player joins the coalition either sooner or later.
This was criticized early on, e.g., by \citeN[p.~255]{Luce/Raiffa:1957}. It is still not widely known that the roll call interpretation of the Shapley value  extends considerably beyond uniform ``yes'' votes.

Namely, a voter can also be decisive for rejecting a proposal by voting ``no'' and being first to ensure that the required majority cannot be met.
In general, we say player~$i$ is \emph{pivotal} in a given voting sequence if the collective decision may still go either way before $i$'s vote but becomes fully determined by it.
Already \citeANP{Mann/Shapley:1960} (\citeyearNP[p.~4]{Mann/Shapley:1960}; \citeyearNP[p.~153]{Mann/Shapley:1964}) 
observed that player~$i$'s SSI equals $i$'s pivot probability 
if all players vote in a mutually independent way with a common probability $x\in[0,1]$ for ``yes'', not just when $x=1$ or $0$. This was first explicitly proven in \citeN{Felsenthal/Machover:1996rollcall}.

But $\varphi_i(v)$'s roll call interpretation applies even more generally: it is sufficient that players' votes are \emph{exchangeable}, so possibly dependent. This can be deduced from combinatorial results by \citeN[Prop.~4]{Hu:2006}. We give a short non-combinatorial proof here. Our main objective, however, is to show that exchangeability is necessary, too:
$i$'s Shapley value equals its pivot probability in roll call votes with random order \emph{if and only if} players' 
cooperation decisions are exchangeable.

A characterization of when pivotality in role calls reduces to the Shapley value is of interest beyond committee decisions:
binary threshold structures similar to voting appear in diverse contexts. Think, e.g., of dichotomous stability assessments in which loans that are
either performing or non-performing play the role of votes and exceeding a given quota of non-performing loans reflects insolvency. 
And if the usual definition of $i$'s marginal contribution is extended to reflect also the reduction of creatable surplus if $i$ refuses to cooperate, then the roll call interpretation of the Shapley value extends to general coalitional games without full cooperation too. 

\section{Preliminaries}
\label{sec:preliminaries}
\noindent
Consider a set $N=\{1,\dots,n\}$ of $n>0$ players.
A \emph{coalitional game} $v\colon 2^N\to\mathbb{R}$ with $v(\varnothing)=0$ maps each \emph{coalition} $S\subseteq N$ of cooperating
players to a real number, typically interpreted as a surplus that increases from zero to $v(N)$ as more players cooperate.
In voting applications, $i\in S$ reflects a ``yes'' vote by player~$i$. Then the focus is on
\emph{simple (voting) games} with $v(S)\in\{0,1\}$: $v(S)=1$ identifies passage of a proposal, $v(\varnothing)=0$, $v(N)=1$, and $S\subseteq T \Rightarrow v(S)\le v(T)$. Simple games $u_T$ defined by $u_T(S)=1 \Leftrightarrow T\subseteq S$ for given $\varnothing \neq T\subseteq N$ are called \emph{unanimity games} and form a basis of the vector space of coalitional games.

\emph{Values} are operators that map coalitional games to $\mathbb{R}^n$ and thereby suggest an allocation of  $v(N)$,  indicate the distribution of voting power, etc.
A value $\psi$ is called \emph{linear} if  $\psi(\alpha\cdot u+\beta\cdot v)=\alpha\cdot\psi(u)+\beta\cdot\psi(v)$ for all constants $\alpha,\beta\in\mathbb{R}$ and all coalitional
games $u,v$ on the same set $N$ of agents, where $\left(\alpha\cdot u+\beta\cdot v\right)(S)= \alpha\cdot u(S)+\beta\cdot v(S)$ for all $S\subseteq N$.
$\psi$ is called \emph{efficient} if $\sum_{i\in N} \psi_i(v)=v(N)$.
A player $i\in N$ satisfying $v(S)=v(S\cup\{i\})$ for all $S\subseteq N\setminus\{i\}$ is called \emph{null}.
If $\psi_i(v)=0$ whenever $i$ is a null player in $v$, then $\psi$ satisfies the \emph{null player property}.
Players $i,j\in N$ with $v(S\cup\{i\})=v(S\cup\{j\})$
for all $S\subseteq N\setminus\{i,j\}$ are called \emph{equivalent}. $\psi$ is \emph{symmetric} if
$\psi_i(v)=\psi_j(v)$ whenever $i,j\in N$ are equivalent in $v$.

Denote the set of all permutations of $N$ by $\SG$ and let $P_i^\pi$ be the set of all agents that precede $i$ in order $\pi\in \SG$. Then the \emph{Shapley value} $\varphi$ is defined by
\begin{equation}
  \label{eq_roll_call_sv}
  \varphi_i(v)=\frac{1}{n!}\cdot\sum_{\pi\in \SG} \left[ v\!\left(P_i^\pi\cup\{i\}\right)-v\!\left(P_i^\pi\right)\right]
\ \text{ for all }i\in N.
\end{equation}
This can also be written and more efficiently be computed as
\begin{equation}
  \label{eq_simplification_sv}
  \varphi_i(v)=\sum_{S\subseteq N\setminus\{i\}} \frac{|S|!\cdot(n-|S|-1)!}{n!}\cdot \left[v(S\cup \{i\})-v(S)\right], 
\end{equation}
i.e., by summing only over $2^{n-1}$ coalitions instead of $n!$ permutations. 
\citeN{Shapley:1953} proved that $\varphi$ is the unique value that satisfies efficiency, linearity, symmetry, and the null player property.

Shapley also gave eq.~(\ref{eq_roll_call_sv}) a roll call interpretation: assume that all players consent to cooperate one after the other. Given an ordering $\pi\in \SG$, player~$i$'s effect on the joint surplus at the time when $i$ decides is $v\!\left(P_i^\pi\cup\{i\}\right)-v\!\left(P_i^\pi\right)$. Considering
all orderings to be equiprobable and taking expectations gives eq.~(\ref{eq_roll_call_sv}).

\citeN[p.~789]{Shapley/Shubik:1954} mentioned for simple games that one can equivalently arrive at $\varphi_i(v)$ assuming all players vote ``no''. If a player decides not to cooperate in a coalitional game, then formation of the grand coalition $N$ is blocked; the player rescinds some surplus that might potentially be created. At the time of choosing not to cooperate, the size of this destructive effect of player~$i$'s non-cooperation is
\begin{equation}
  v\!\left(N\setminus P_i^\pi\right)-
  v\!\left(N\setminus\left(P_i^\pi\cup\{i\}\right)\right)=
  v^*\!\left(P_i^\pi\cup\{i\}\right)-v^*\!\left(P_i^\pi\right),
\end{equation}
where $v^*(S):=v(N)-v(N\setminus S)$ for all $S\subseteq N$ defines the \emph{dual game} of $v$ and $\varphi(v^*)=\varphi(v)$. 

Allowing cooperation (``yes'') by some players and non-cooperation (``no'') by others
gives rise to a \emph{generalized roll call model}
that was introduced by \citeANP{Mann/Shapley:1960} (\citeyearNP[p.~4]{Mann/Shapley:1960}; \citeyearNP[p.~153]{Mann/Shapley:1964}) and 
taken up by \citeN{Felsenthal/Machover:1996rollcall}:
an ordering $\pi$ of players is determined; each player~$i\in N$ is called in order; when called, $i$ decides either to cooperate or not.
Denoting the resulting final sets of cooperators or supporters of a motion by $S$ and the non-cooperators by $\barS:=N \setminus S$, the actual surplus created is $v(S)$; the potential surplus rescinded is $v^*(\barS)=v(N)-v(S)$.
A particular instance of a roll call will be referred to as $\R=(\pi,S)$ for $\pi\in \SG$ and $S\in 2^N$.

To assess the effect of a given player~$i$ in this process of (non-)creation in game $v$, let  $\mathcal{Y}(\R,i)$ denote the set of cooperative players $j\in S$ that precede player~$i$.
Similarly, let $\mathcal{N}(\R,i)$ collect all uncooperative players $j\in \barS$ that precede $i$. We can then define the \emph{marginal contribution} of player~$i$ in roll call $\R$ for game $v$ as
\begin{equation} \label{eq:M-rollcall}
  M(v,\R,i)=\left\{
  \begin{array}{rl}
    v(\mathcal{Y}(\R,i)\cup\{i\})-v(\mathcal{Y}(\R,i)) & \text{if } i\in S,\\
    v^*(\mathcal{N}(\R,i)\cup\{i\})-v^*(\mathcal{N}(\R,i)) & \text{if } i\in \barS.
  \end{array}
  \right.
\end{equation}
For a simple game $v$, $M(v,\R,i)\in\{0,1\}$ and $M(v,\R,i)=1$ if and only if player~$i$ is pivotal in $\R$: fate of a given proposal is still open before $i$'s vote but sealed by $i$'s decision.

Player~$i$'s overall effect or power in game $v$ can be captured by computing its expected marginal contribution for an appropriate distribution over roll calls. We stay in line with eq.~(\ref{eq_roll_call_sv}) by presuming that orderings are drawn independently from the uniform distribution on $\mathcal{S}_n$. However, we define value $\varphi^p$ by
\begin{equation}
  \label{eq_gen_value}
  \varphi_i^p(v)=\frac{1}{n!}\sum_{\pi\in \SG}\sum_{S\in 2^N} p(S)\cdot M(v,(\pi,S),i) \ \text{ for }i\in N
\end{equation}
for an arbitrary 
probability distribution $p$ on $2^N$, i.e., requiring only $p(S)\ge 0$ for all $S\in 2^N$ and $\sum_{S\in 2^N} p(S)=1$. Cooperation of players thus neither needs to be complete with $p(N)=1$, nor independent with $p(S)=\prod_{i\in S}x_i \prod_{i\not\in S}(1-x_i)$ for $x_i\in [0,1]$.

\section{Results}

\label{sec_main}
\noindent
\begin{proposition}
  \label{prop_1}
  Value $\varphi^p$ is linear, efficient, and satisfies the null player property for every probability distribution $p$.
\end{proposition}

\noindent \textit{Proof}
  The null player property is obvious from the definition. Linearity follows from recalling that $v^*(\mathcal{N}(\R,i)\cup\{i\})-v^*(\mathcal{N}(\R,i))=v(N\setminus \mathcal{N}(\R,i))-v(N\setminus (\mathcal{N}(\R,i)\cup\{i\}))$. So
  $\varphi^p$ is a linear combination of terms that are linear in $v$.
  For efficiency, first observe that
\begin{equation}
    \sum_{i=1}^n M(v,\R,i) = v(S)-v(\varnothing)+v^*(\barS)-v^*(\varnothing)=v(N)-v(\varnothing)=v(N)
\end{equation}
  for any $\R\in \SG\times 2^N$ given the telescope sum behavior of $\sum_{i=1}^n M(v,\R,i)$. Second, $\left|\SG\right|=n!$ and $\sum_{S\in 2^N} p(S)=1$.
\hfill{$\square$}
\bigskip

Random variables $X_1, \ldots, X_n$ are called \emph{exchangeable} or \emph{symmetrically dependent} if the $n!$ permutations $(X_{k_1}, \ldots, X_{k_n})$ all have the same $n$-dimensional probability distribution (see, e.g., \citeNP[sec.~7.4]{Feller:1971}). Applied to votes or binary cooperation choices, which $\varphi^p$ treats as random variables, this is equivalent to $p(S)=p(S')$ whenever $|S|=|S'|$, i.e., the probability of a particular partition of $N$ into cooperators $S$ and non-cooperators $\barS$ depends only on the number of (non-)cooperators rather than their identities.

\begin{proposition}
	\label{prop_2}
	If players' cooperation choices are exchangeable under $p$ then  $\varphi^p$ is symmetric.
\end{proposition}

\noindent \textit{Proof}
Let $\kappa$ denote the permutation that swaps players $i$ and $j$
and define $\kappa(\R)=(\kappa(\pi),\kappa(S))$ for any given roll call $\R=(\pi,d)$. If $i$ and $j$ are equivalent then $M(v,\R,i)=M(v,\kappa(\R),j)$ for all $\R$. Exchangeability implies $p(S)=p(\kappa(S))$.
Hence
\begin{align}
\varphi_i^p(v)&=\frac{1}{n!}\sum_{\pi\in \SG}\sum_{S\in 2^N} p(S)\cdot M(v,(\pi,S),i)  
 =\frac{1}{n!}\sum_{\pi\in \SG}\sum_{S\in 2^N} p(\kappa(S))\cdot M(v,\kappa(\pi,S),j)  \\
& 
=\frac{1}{n!}\sum_{\pi'\in \SG}\sum_{S' \in 2^N} p(S')\cdot M(v,(\pi',S'),j) =  \varphi_j^p(v). \notag
\end{align}

\vspace{-1cm} \hfill{$\square$}

\bigskip

\noindent Proposition~\ref{prop_1} and the characterization of $\varphi$ by \citeN{Shapley:1953} then give 
the generalization of \citeANP{Felsenthal/Machover:1996rollcall}'s \citeyear{Felsenthal/Machover:1996rollcall} result by
\citeN[Prop.~4]{Hu:2006} as an immediate corollary: if players' cooperation choices are exchangeable under $p$ then $\varphi^p(v)=\varphi(v)$ for all coalitional games $v$.\footnote{A special case of Hu's result, namely  
	$\varphi^p(v)=\varphi(v)$ if $p(S)=2^{-n}$ for all $S\subseteq N$, 
	was 
	published by \citeN{Bernardi/Freixas:2018}. A combinatorial proof of Hu's result also is contained in \citeN{Kurz:2016arxiv}.}
We here show that the converse holds too: 
\begin{proposition}
	\label{prop_3}
	If $\varphi^p$ is symmetric then players' cooperation choices are exchangeable under $p$.
\end{proposition}

\noindent
\textit{Proof} 
We need to prove that $|S|=|S'|\Rightarrow p(S)=p(S')$ if $\varphi^p$ is symmetric. This is satisfied trivially if $n=1$, $S=\varnothing$, or $S=N$ since then $S=S'$. So consider $n\ge 2$, $S\in 2^N\setminus \{\varnothing, N\}$ and $S\neq S'$.
The symmetric difference $S\Delta S':=\{i\in N: i\notin S\cap S'\}$ contains between 2 and  $2\cdot |S|$ members. But there always exists a finite path $(X_1, \ldots, X_r)$ with $X_1=S$ and $X_r=S'$ such that $X_l$ and $X_{l+1}$ differ by just one player $i\in X_l$ being replaced by some $j\notin X_l$, i.e., $X_l \Delta X_{l+1}=\{i,j\}$. To prove the claim, it therefore suffices to show that symmetry of $\varphi^p$ implies $p(X\cup \{i\})=p(X\cup \{j\})$ for every set $X\subseteq N\setminus \{i,j\}$ and $i\neq j\in N$.

Fix any such set $X\subseteq N\setminus \{i,j\}$ and let us consider the game
\begin{equation}
v_X=\sum_{\{i,j\}\subseteq T\subseteq N} \lambda_{T,X}\cdot u_T,
\end{equation}
where $\lambda_{T,X}=M^{-1}_{X,(T\setminus\{i,j\})}$ invokes the inverse matrix $M^{-1}$ specified as follows:
\begin{lemma}
	\label{lemma_linalg}
	Let $M$ be the $2^m\times 2^m$ matrix $M$ defined by $M_{R,S}=\frac{1}{1+|R\setminus S|}$ for
	all $R,S\subseteq G$, where $G$ is a set of cardinality $m$. 
	Then $M$'s inverse $M^{-1}$ is given by
	\begin{equation}
	M^{-1}_{R,S}={{m+1}\choose {m+|R\setminus S|}}\cdot (-1)^{|S\Delta R|}.
	\end{equation}
	\end{lemma}
\medskip

\noindent Proof of Lemma~\ref{lemma_linalg} is provided in the appendix. Game $v_X$ is chosen such that its coordinates $\lambda_{T,X}$ in the unanimity game basis $\{u_T\}$ of the space of coalitional games satisfy
\begin{equation}
\label{eq_vanish2}
\sum_{\{i,j\}\subseteq T\subseteq N} \lambda_{T,X}\cdot \frac{1}{1+|(T\setminus \{i,j\})\setminus S |}
=M^{-1}_{X,(T\setminus\{i,j\})}\cdot M_{(T\setminus\{i,j\}),S}
=\begin{cases}0, & S\neq X, \\1, &S=X\end{cases}
\end{equation}
for any given set $S\subseteq N\setminus \{i,j\}$. This follows from Lemma~\ref{lemma_linalg} using $G=N\setminus \{i,j\}$ and  $R=T\setminus\{i,j\}$. Since $v_X$ is a linear combination of unanimity games $u_T$ in that $i$ and $j$ are equivalent, they are equivalent in $v_X$.

For any unanimity game $u_T$ with $i,j\in T$, value $\varphi^p_i(u_T)$ captures pivotality of $i$  in two situations:
if $i \in S$, then $i$ is pivotal in roll call $\R=(\pi,S)$ iff all players in $T$ vote ``yes'' (so $T\subseteq S$) and $i$ is the last member of $T$ to be called;
if $i\notin S$, then player~$i$ is pivotal iff $i$ is the first member of $T$ to be called. 
If we divide the latter roll calls with $i\notin S$ according to whether $j\in S$ or $j\notin S$, we have
\begin{equation}
\varphi^p_i(u_T)=\sum_{T\subseteq S\subseteq N} \frac{1}{|T|}\cdot p(S)+
\sum_{S\subseteq N\setminus \{i,j\}} \frac{1}{|T\setminus S|}\cdot p(S)+\sum_{S\subseteq N\setminus \{i,j\}} \frac{1}{|T\setminus S|-1}\cdot p(S\cup\{j\}). \label{eq_vanish}
\end{equation}

Symmetry of $\varphi^p$ and linearity (Prop.~\ref{prop_1}) imply
\begin{equation}\label{eq:equating_phi_i_and_phi_j}
\varphi^p_i(v_X)=\sum_{\{i,j\}\subseteq T\subseteq N} \lambda_{T,X}\cdot \varphi^p_i(u_T) =
\sum_{\{i,j\}\subseteq T\subseteq N} \lambda_{T,X}\cdot \varphi^p_j(u_T)=\varphi^p_j(v_X).
\end{equation}
The expressions for $\varphi^p_j(u_T)$ analogous to eq.~(\ref{eq_vanish}) involve identical first and second summands. Cancelling these in eq.~(\ref{eq:equating_phi_i_and_phi_j}) yields
\begin{equation}
\sum_{\{i,j\}\subseteq T\subseteq N} \sum_{ S\subseteq N\setminus \{i,j\}} \lambda_{T,X} \cdot \frac{1}{|T\setminus S|-1}\cdot p(S\cup\{j\})=\sum_{\{i,j\}\subseteq T\subseteq N} \sum_{S\subseteq N\setminus \{i,j\}} \lambda_{T,X} \cdot \frac{1}{|T\setminus S|-1}\cdot p(S\cup\{i\}).
\end{equation}
Changing the order of summation, noting $|T\setminus S|-1=1+|T\setminus S\setminus \{i,j\}|$ if $\{i,j\}\subseteq T$, and invoking eq.~(\ref{eq_vanish2}) reduces this to $p(X\cup\{j\})=p(X\cup\{i\})$. This proves the claim.
\hfill{$\square$}
\bigskip

As a direct corollary to Propositions~\ref{prop_1}--\ref{prop_3} and \citeN{Shapley:1953}, we obtain a full characterization of when the Shapley value has a roll call interpretation:
\begin{theorem}
	\label{main_theorem}
$\varphi^p(v)=\varphi(v)$ for all coalitional games $v$ if and only if players' cooperation choices are exchangeable under $p$.
\end{theorem}

\noindent It is clear from the proof of Proposition~\ref{prop_3} that coincidence of $\varphi^p$ and $\varphi$ in Theorem~\ref{main_theorem} could be restricted to any subclass of games which includes basis $\{u_T\}$, such as simple games.

The marginal contribution of player~$i$ in roll call $\R$ defined in eq.~(\ref{eq:M-rollcall}) is key to interpreting $\varphi_i^p(v)$. It is possible to give rather general economic meaning to it in terms of a player's effect on both the created and the rescinded surplus associated with formation of a coalition $S\subseteq N$. 

To us the roll call interpretation of the Shapley value is most appealing for simple voting games. Then, for a given joint distribution $p$ that describes the ``yes''-or-``no'' inclinations of voters and considering uniformly random sequences of players being called,  $\varphi_i^p(v)$ is the probability of player~$i$ being pivotal: either conclusively passing the proposal or putting the final nail in its coffin.
Pivot probabilities are widely applied in order to assess how given  voting rules translate into a distribution of voting power in various decision bodies (e.g., shareholder meetings, the US~Electoral College, EU~Council of Ministers, IMF~Board of Directors, etc.; cf.\ \citeNP{Napel:2018}). They 
also are of interest in other environments that involve binary variables, such as in reliability analysis of components or factors whose functionality is critical to a technical system or success of a project.

Theorem~\ref{main_theorem} characterizes all scenarios such that the Shapley value captures players' pivot probabilities. These include 
the textbook case of all players voting ``yes'' as well as independent votes with a probability $x\in [0,1]$ for ``yes''. But they go considerably beyond: the Shapley value equals pivot probabilities in roll calls if and only if votes are exchangeable.


\appendix
\section*{Appendix}
\subsection*{Proof of Lemma~\ref{lemma_linalg}}

The proof 
draws on the following two combinatorial claims with $n\in \{0,1,2,\ldots\}$:
\begin{claim}
	\label{claim_binomial_sum1}
\begin{equation*}
\sum_{k=0}^{n}{n\choose k}\cdot (-1)^k=\begin{cases} 1 &\text{ if } n=0, \\ 0 & \text{ if } n\ge 1. \end{cases} 
\end{equation*}
\end{claim}	
\noindent 
\textit{Proof} This follows from $\sum_{k=0}^{0}{0\choose k}\cdot (-1)^k={0\choose 0}=1$ and the binomial theorem, i.e.,
$	0=(1-1)^n=\sum_{k=0}^{n}{n\choose k}\cdot (-1)^k$  for $n\ge 1$. \hfill{$\square$}

\begin{claim}	\ ~
		\label{claim_binomial_sum2}
\begin{enumerate}
\item[\textnormal{(a)}]	
\begin{equation*}
 \sum\limits_{k=0}^n {n \choose k}\cdot\frac{(-1)^k}{k+1}=\frac{1}{n+1},
\end{equation*}
 \item[\textnormal{(b)}]
\begin{equation*}
 \sum\limits_{k=0}^n {n \choose k}\cdot\frac{(-1)^k}{k+x}=\frac{n!}{\prod_{k=0}^{n} (x+k)} \quad \text{ for all } x\in(0,\infty),
\end{equation*}

\item[\textnormal{(c)}]
\begin{equation*}
\sum\limits_{k=0}^n {n \choose k}\cdot\frac{(-1)^k}{k+1+x}=\frac{n!}{\prod_{k=0}^{n} (1+x+k)} \quad \text{ for all }x\in(-1,\infty).
\end{equation*}
\end{enumerate}
\end{claim}
\textit{Proof} Consider the following polynomial of degree at most $n$
\begin{equation}
	f(x)=\sum_{k=0}^n {n\choose k}\cdot(-1)^k \cdot\prod_{0\le j\le n\,:\, j\neq k} (x+1+j). \label{eq:polynomial}
\end{equation}
For every $i\in \{0,1,\ldots, n\}$ we have
\begin{equation}
	f(-i-1)={n\choose i}\cdot(-1)^i \cdot\prod_{0\le j\le n\,:\, j\neq i} (x+1+j)
	={n\choose i}\cdot(-1)^i \cdot (-1)^i i!\cdot (n-i)!=n!
\end{equation}
since each product in eq.~(\ref{eq:polynomial}) contains one factor $(j-i)=0$ when $k\neq i$.
A polynomial of degree at most $n$ that equals $n!$ for $n+1$ distinct $x$ must be constant; so $f(x)=n!$. Part~(c) then follows from division by $\prod_{k=0}^{n} (1+x+k)$.
Setting $x=0$ in part~(c) yields (a). Part~(b) follows from (c) by a transformation of variable. \hfill{$\square$}

\smallskip

Now consider matrices $M$ and $M^{-1}$ with
\begin{equation}
M_{R,S}=\frac{1}{1+|R\setminus S|}
\quad\quad \text{ and } \quad\quad
M^{-1}_{R,S}={{m+1}\choose {m+|R\setminus S|}}\cdot (-1)^{|S\Delta R|}
\end{equation}
for all $R, S\subseteq G=\{1,\dots,m\}$
and let us show that $\left(M\cdot M^{-1}\right)_{R,S}=\sum_{U\subseteq G} M_{R,U}\cdot M^{-1}_{U,S}$ equals the $2^m\times 2^m$-identity matrix. All terms involving $M^{-1}_{U,S}$ with $|U\setminus S|\ge 2$ vanish since ${m \choose k}=0$ for $k>m$. The remaining terms either involve $U\subseteq S$ with $|U\setminus S|=0$ (implying ${{m+1}\choose {m+|U\setminus S|}}=m+1$); or $U$ such that $|U\setminus S|=1$ implying ${{m+1}\choose {m+|U\setminus S|}}=1$ and $|U\Delta S|=1+|S\setminus U|$. So
	\begin{eqnarray}
		\left(M\cdot M^{-1}\right)_{R,S}
		&=& \sum_{U\subseteq S} \frac{m+1}{1+|R\setminus U|}\cdot (-1)^{|S\setminus U|} \notag
		+\sum_{U\subseteq G\,:\, |U\setminus S|=1} \frac{-1}{1+|R\setminus U|}\cdot (-1)^{|S\setminus U|}\\
		&=& \sum_{U\subseteq S} \frac{m+1}{1+|R\setminus U|}\cdot (-1)^{|S\setminus U|}
		+\sum_{U\subseteq S}\sum_{l\in G\setminus S} \frac{-1}{1+|R\setminus (U\cup\{l\})|}\cdot (-1)^{|S\setminus U|} \label{eq:MtimesM-1}
	\end{eqnarray}
	Let us use the abbreviations $a=|R\cap S|$ and $b=|S\setminus R|$, so that $a+b=|S|$. With this we compute
	\begin{eqnarray}
		&&\sum_{U\subseteq S} \frac{m+1}{1+|R\setminus U|}\cdot (-1)^{|S\setminus U|} \notag \\
		&=& \sum_{i=0}^{a}\sum_{j=0}^{b} {a\choose i}{b\choose j} \cdot\frac{m+1}{1+|R|-i}\cdot(-1)^{|S|-i-j} \notag  \\
		&=& \sum_{i=0}^{a} {a\choose i} \cdot\frac{m+1}{1+|R|-i}\cdot(-1)^{|S|-i}\cdot \left(\sum_{j=0}^{b} {b\choose j}\cdot (-1)^j\right) \label{eq:l_in_U}
	\end{eqnarray}
	and
	\begin{eqnarray}
		&&\sum_{U\subseteq S}\sum_{l\in G\setminus S} \frac{-1}{1+|R\setminus (U\cup\{l\})|}\cdot (-1)^{|S\setminus U|} \notag \\
		&=& \sum_{l\in G\setminus S}\sum_{i=0}^{a}\sum_{j=0}^{b} {a\choose i}{b\choose j} \cdot\frac{-1}{1+|R\setminus\{l\}|-i}\cdot(-1)^{|S|-i-j} \notag \\
		&=& -\sum_{l\in G\setminus S}\sum_{i=0}^{a}{a\choose i} \cdot\frac{1}{1+|R\setminus\{l\}|-i}\cdot(-1)^{|S|-i}\cdot\left(\sum_{j=0}^{b}{b\choose j}\cdot (-1)^j\right).\label{eq:l_notin_U}
	\end{eqnarray}
If $b>0$, corresponding to $S\not\subseteq  R$, Claim~\ref{claim_binomial_sum1} implies $\left(M\cdot M^{-1}\right)_{R,S}=0$

It remains to consider $b=0$, corresponding to $S\subseteq R$. Claim~\ref{claim_binomial_sum1} then implies $\left(\sum_{j=0}^{b}{b\choose j}\cdot (-1)^j\right)=1$, which simplifies expressions (\ref{eq:l_in_U}) and (\ref{eq:l_notin_U}).  The case  $|U\setminus S|=1$ captured by (\ref{eq:l_notin_U}) splits into $x:=|R\setminus S|$ subcases where $l\in G\setminus S$ is member of $R$, and $m-|S|-x$ subcases where $l$ is neither member of $S$ nor of $R$.
Noting, moreover, that $|R|=x+|S|$ we can use this case distinction to  write (\ref{eq:MtimesM-1}) as
	\begin{eqnarray}
\left(M\cdot M^{-1}\right)_{R,S}
		&=& \sum_{i=0}^{|S|} {{|S|}\choose i} \cdot\frac{(m+1)\cdot (-1)^{|S|-i}}{1+x+|S|-i}
		-\sum_{i=0}^{|S|} {{|S|}\choose i} \cdot\frac{(m-|S|-x)\cdot (-1)^{|S|-i}}{1+x+|S|-i} \notag \\
		&&-\sum_{i=0}^{|S|} {{|S|}\choose i} \cdot\frac{x\cdot (-1)^{|S|-i}}{x+|S|-i} \notag \\
		&=& \sum_{i=0}^{|S|} {{|S|}\choose i} \cdot\frac{(|S|+x+1)\cdot (-1)^{|S|-i}}{1+x+|S|-i}
		-\sum_{i=0}^{|S|} {{|S|}\choose i} \cdot\frac{x\cdot (-1)^{|S|-i}}{x+|S|-i} \notag \\
		&=& (|S|+x+1)\cdot \sum_{k=0}^{|S|} {{|S|}\choose k}\cdot \frac{ (-1)^{k}}{1+x+k}
		-x\cdot \sum_{k=0}^{|S|} {{|S|}\choose k}\cdot\frac{ (-1)^{k}}{x+k}.
	\end{eqnarray}
	For $x=0$, i.e., $S=R$, Claim~\ref{claim_binomial_sum2}(a) then gives $\left(M\cdot M^{-1}\right)_{R,R}=1$.
	For $x>0$, i.e., $S\subsetneq R$, Claims~\ref{claim_binomial_sum2}(b) and \ref{claim_binomial_sum2}(c) give
\begin{equation}
	\left(M\cdot M^{-1}\right)_{R,S}=
	(|S|+x+1)\cdot \frac{m!}{\prod_{k=0}^{|S|} (1+x+k)}
	-x\cdot \frac{m!}{\prod_{k=0}^{|S|} (x+k)}=0.
\end{equation}

\noindent In summary, we have
\begin{equation}
	\left(M\cdot M^{-1}\right)_{R,S}=\begin{cases}
	1 & \text{if }R=S, \\
	0 &\text{otherwise.} 
	\end{cases}
\end{equation}
\smallskip

\vspace{-1cm} \hfill{$\square$}

\subsection*{Acknowledgements}
\noindent
We thank Pradeep~Dubey, Piero~La~Mura, and all other participants of the 5th~Workshop on Cooperative Game Theory in Business Practice, Leipzig, 
for helpful feedback. We also acknowledge constructive comments that we received from an anonymous referee. The usual caveat applies.

\setlength{\labelsep}{-0.2cm}


\begin{thebibliography}{}
	\bibitem[\protect\citeauthoryear{Bernardi and Freixas}{Bernardi and Freixas}{2018}]{Bernardi/Freixas:2018}
	Bernardi, G. and J.~Freixas (2018).
	\newblock The {S}hapley value analyzed under the {F}elsenthal and {M}achover bargaining model.
	\newblock {\em Public Choice\/}~{\em 176\/}(3-4), 557--565.
	
	
	\bibitem[\protect\citeauthoryear{Feller}{Feller}{1971}]{Feller:1971}
	Feller, W. (1971).
	\newblock {\em An Introduction to Probability Theory and Its Applications},
	Volume~2.
	\newblock New York, NY: Wiley.
	
	\bibitem[\protect\citeauthoryear{Felsenthal and Machover}{Felsenthal and
		Machover}{1996}]{Felsenthal/Machover:1996rollcall}
	Felsenthal, D.~S. and M.~Machover (1996).
	\newblock Alternative forms of the {S}hapley value and the {S}hapley-{S}hubik
	index.
	\newblock {\em Public Choice\/}~{\em 87\/}(3-4), 315--318.
	
	\bibitem[\protect\citeauthoryear{Hu}{Hu}{2006}]{Hu:2006}
	Hu, X. (2006).
	\newblock An asymmetric {S}hapley--{S}hubik power index.
	\newblock {\em International Journal of Game Theory\/}~{\em 34\/}(2), 229--240.
	
   \bibitem[\protect\citeauthoryear{Kurz}{Kurz}{2016}]{Kurz:2016arxiv}
	Kurz, S. (2016).
	\newblock Generalized roll-call model for the Shapley-Shubik index.
	\newblock arXiv preprint 1602.04331. 
	
	
	\bibitem[\protect\citeauthoryear{Luce and Raiffa}{Luce and
		Raiffa}{1957}]{Luce/Raiffa:1957}
	Luce, R.~D. and H.~Raiffa (1957).
	\newblock {\em Games and Decisions}.
	\newblock New York: Wiley.
	
		\bibitem[\protect\citeauthoryear{Mann and Shapley}{Mann and
		Shapley}{1960}]{Mann/Shapley:1960}
	Mann, I. and L.~S. Shapley (1960).
	\newblock Values of large games, IV: evaluating the Electoral College by Montecarlo techniques.
	\newblock Research Memorandum 2651. 
\newblock	RAND Corporation, Santa Monica, CA.
	
	\bibitem[\protect\citeauthoryear{Mann and Shapley}{Mann and
		Shapley}{1964}]{Mann/Shapley:1964}
	Mann, I. and L.~S. Shapley (1964).
	\newblock The a priori voting strength of the Electoral College.
	\newblock In M.~Shubik (Ed.), {\em Game Theory and Related Approaches to Social
		Behavior}, pp.\  151--164. Huntington, NY: Robert E.\ Krieger Publishing.
	
	\bibitem[\protect\citeauthoryear{Napel}{Napel}{2019}]{Napel:2018}
	Napel, S. (2019).
	\newblock Voting power.
	\newblock In R.~Congleton, B.~Grofman, and S.~Voigt (Eds.), {\em Oxford
		Handbook of Public Choice}, Volume~I, Chapter 6. Oxford: Oxford University Press.
	
	\bibitem[\protect\citeauthoryear{Shapley}{Shapley}{1953}]{Shapley:1953}
	Shapley, L.~S. (1953).
	\newblock A value for $n$-person games.
	\newblock In H.~W. Kuhn and A.~W. Tucker (Eds.), {\em Contributions to the
		Theory of Games}, Volume~II, pp.\  307--317. Princeton, NJ: Princeton
	University Press.
	
	\bibitem[\protect\citeauthoryear{Shapley and Shubik}{Shapley and
		Shubik}{1954}]{Shapley/Shubik:1954}
	Shapley, L.~S. and M.~Shubik (1954).
	\newblock A method for evaluating the distribution of power in a committee
	system.
	\newblock {\em American Political Science Review\/}~{\em 48\/}(3), 787--792.
	
\end{thebibliography}
\newcommand{\noopsort}[1]{}

\end{document}